\documentclass[twocolumn,aps,prl,10pt,showpacs,dvips]{revtex4}
\usepackage{epsfig}
\begin{document}
\title{{\Large Microscopic theory of a phase transition in a critical region:\\
Bose-Einstein condensation in an interacting gas}}

\author{Vitaly V. Kocharovsky$^{a,b,*}$ and Vladimir V. Kocharovsky$^{b,c}$}

\affiliation{$^{a}$Department of Physics and Astronomy, Texas A\&M University, College Station, TX 77843-4242, USA\\
$^{b}$Institute of Applied Physics, Russian Academy of Science,
603950 Nizhny Novgorod, Russia\\
$^{c}$University of Nizhny Novgorod, 23 Gagarin Avenue, Nizhny Novgorod 603950, Russia}

\date{\today}

\begin{abstract}
    We present a microscopic theory of the second order phase transition in an interacting Bose gas that allows one to describe formation of an ordered condensate phase from a disordered phase across an entire critical region continuously. We derive the exact fundamental equations for a condensate wave function and the Green's functions, which are valid both inside and outside the critical region. They are reduced to the usual Gross-Pitaevskii and Beliaev-Popov equations in a low-temperature limit outside the critical region. The theory is readily extendable to other phase transitions, in particular, in the physics of condensed matter and quantum fields.
    
    Published in Physics Letters A {\bf 379}, 466 (2015), http://dx.doi.org/10.1016/j.physleta.2014.10.052.
    
\end{abstract}
\maketitle
%\footnotetext[]{*Corresponding author: Department of Physics and %Astronomy,\\Texas A\&M University, College Station, TX 77843-4242, USA.\\
%E-mail address: vkochar@physics.tamu.edu (V.V. Kocharovsky)}

     The problem of finding a microscopic theory of the second order phase transitions became one of the major problems in theoretical physics in 1950s after an invention of the powerful Green's functions and quantum field theory methods in the many-body physics. These methods include the Feynman's and Matsubara's diagram techniques, Wick's theorem, and Dyson's equation (see \cite{AGD,HohenbergMartin,FetterWalecka,Anderson1984,Shi,PitString,RevModPhys2004,ProukakisTutorial2008,LLV,LL,PatPokr,Fisher1986,Goldenfeld,Vicari2002} and references therein). A major point is a failure of a Landau self-consistent field theory in a critical region, shown by the exact Onsager's solutions. However, all these standard methods turn out to be insufficient to solve the problem: The microscopic theory, which should connect the asymptotics of the ordered and disordered phases across the critical region, has not been found so far even for anyone of the numerous phase transitions.
     
     Here we present such a microscopic theory for a Bose-Einstein condensation (BEC) in an interacting gas. We derive the fundamental equations for the order parameter and Green's functions, which are valid not only in the fully ordered, low-temperature phase, but also in the entire critical region and in the disordered phase. They resolve a fine structure of the system's statistics and thermodynamics near a critical $\lambda$-point. It becomes possible due to the newly developed methods of (a) the nonpolynomial averages and contraction superoperators \cite{KochLasPhys2007,KochJMO2007}, (b) the partial difference (recurrence) equations \cite {PDE-Cheng,DE-Agarwal,DE-Elaydi} (a discrete analog of the partial differential equations) for superoperators, and (c) a characteristic function and cumulant analysis for a joint distribution of the noncommutative observables. They allow us to take into account 
     
     (I) the constraints in a many-body Hilbert space, which are the integrals of motion prescribed by a broken symmetry in virtue of a Noether's theorem, and constraint-cutoff mechanism, responsible for the very existence of a phase transition and its nonanalytical features, \cite{Anderson1984,PRA2010,PRA2014}
     
     (II) an insufficiency of a grand-canonical-ensemble approximation, which is incorrect in the critical region \cite{HohenbergMartin,ProukakisTutorial2008} because of averaging over the systems with different numbers of particles, both below and above the critical point, i.e., over the condensed and noncondensed systems at the same time, that implies an error on the order of $100\%$ for any critical function,
     
     (III) a necessity to solve the problem for a finite system with a mesoscopic (i.e., large, but finite) number of particles $N$ in order to calculate correctly an anomalously large contribution of the lowest energy levels to the critical fluctuations and to avoid the infrared divergences of the standard thermodynamic-limit approach \cite{Shi,PitString,RevModPhys2004,ProukakisTutorial2008,LLV,LL,PatPokr} as well as to resolve a fine structure of the $\lambda$-point,
     
     (IV) a fact that in the critical region the Dyson-type closed equations do not exist for true Green's functions, but do exist for the partial 1- and 2-contraction superoperators, which reproduce themselves under a contraction.
     
     The problem of the critical region and mesoscopic effects is directly related to numerous modern experiments and numerical studies on the BEC of a trapped gas (including BEC on a chip), where $N\sim 10^{2}-10^{7}$, (see, for example, \cite{BrownTwissOnBECThresholdNaturePhys2012,Hadzibabic2013NaturePhys,Hadzibabic2013,EsslingerCriticalBECScience2007,TwinAtomBeamNaturePhys2011,Dalibard2012NaturePhys,Dalibard2010,RaizenTrapControl,Blume,BECinterferometerOnChip2013,EntanglementOnChipNature2010,chipBECNature2001}) and superfluidity of $^4$He in nanodroplets ($N\sim 10^{8}-10^{11}$) \cite{HeDroplets} and porous glasses \cite{Reppy}.
     
     A standard renormalization-group theory of the phase transitions, developed since 1960s, \cite{LLV,LL,PatPokr,Fisher1986,Goldenfeld,Vicari2002} does not solve the problem since it is phenomenological and employs a grand-canonical-ensemble approximation that neglects the constrains in a many-body Hilbert space, imposed by a broken symmetry. Moreover, it mostly deals with the thermodynamic-limit quantities and focuses on a few first terms in an asymptotic expansion at a critical-point vicinity, related mainly to the critical exponents. 
     
     All issues (I)-(IV) are usually missed in a BEC theory, but, along with the many-body and nonlinear nature of BEC, have to be included in the microscopic theory to make it right. An absence of an acceptable diagram technique in the canonical ensemble, which has to be used instead of the unconstrained, grand canonical one in view of (I) and (II) but was not found despite many attempts \cite{KwokWoo}, partly explains why the microscopic theory of the phase transitions was not found for so long time. 
     
     BEC occurs only if a number of particles is conserved. For example, a system of photons in equilibrium inside a black-box cavity does not show BEC for it lacks a gauge symmetry due to absorption in the cavity's walls and, hence, the number of photons is not its integral of motion. 

$ $

\noindent {\bf 1. Symmetry-constrained Hilbert space and Hamiltonian}

\noindent An exact reduction of a many-body Hilbert space due to a symmetry constraint is what one has to start with. We limit an unconstrained Fock space $\mathcal{H}^{(0)}$, which spans all Fock states $|n_{\bf k}\rangle$ with integer occupations $n_{\bf k}\in [0,\infty )$, to a physical, canonical-ensemble subspace $\mathcal{H}_N^{(0)}$, restricted by a particle-number constraint $\sum_{\bf k}\hat{n}_{\bf k}=N$, which is an integral of motion in virtue of a global gauge symmetry to be broken in the BEC phase transition. For simplicity's sake, we consider the BEC of $N$ interacting particles with a mass $M$ in a cubic box with volume $V=L^3$ and periodic boundary conditions. The trap's one-particle eigenfunctions $L^{-3/2}e^{i{\bf kr}}$ and energies $\varepsilon_{\bf k}=\frac{\hbar^2 k^2}{2M}$ are specified by a wave-vector quantum number ${\bf k}=\{ k_x , k_y , k_z \}, k_i = \frac{2\pi}{L}q_i$, where $q_i=0, \pm 1, \dots$ is an integer. The ground state ${\bf k}=0$ is nondegenerate. An occupation operator $\hat{n}_{\bf k}=\hat{a}_{\bf k}^{\dagger }\hat{a}_{\bf k}$ is determined by the creation, $\hat{a}_{\bf k}^{\dagger }$, and annihilation, $\hat{a}_{\bf k}$, operators for a ${\bf k}$-eigenstate and obeys a canonical commutation relation $[\hat{a}_{\bf k}, \hat{a}_{\bf k'}^{\dagger }]=\delta_{{\bf k},{\bf k'}}$, where $\delta_{{\bf k},{\bf k'}}$ is a Kronecker's delta. 

    That reduction is achieved by excluding the ground-state component since its occupation $n_0 = N-n$ for any Fock state is determined by the total excited-states occupation $n=\sum_{{\bf k}\neq 0}n_{\bf k}$. The physics (dynamics, fluctuations, etc.) of the system is determined by creation and annihilation of the canonical-ensemble excitations (not particles!) via operators $\hat{\alpha}_{\bf k}^{'\dagger}=\theta(N-\hat{n})\hat{\alpha}_{\bf k}^{\dagger}$ and $\hat{\alpha}^{'}_{\bf k}=\hat{\alpha}_{\bf k}\theta(N-\hat{n})$, which leave invariant (i.e., do not lead out of) the physical, canonical-ensemble subspace $\mathcal{H}_N^{(0)}$. They are the step-function $\theta(N-\hat{n})$ cutoff \cite {PRA2010,PRA2014} of the Girardeau-Arnowitt operators \cite{GA}, defined via a transition operator $\hat{\alpha}_{\bf k}=\hat{a}_0^{\dagger}(1+N-\hat{n})^{-1/2}\hat{a}_{\bf k}$; $\hat{n}=\sum_{{\bf k}\neq 0}\hat{n}_{\bf k}$. Let us introduce also a Fock space $\mathcal{H}$, which spans only the excited states $|n_{\bf k\neq 0}\rangle$, and the creation and annihilation operators in $\mathcal{H}$, $\hat{\beta}_{\bf k}^{\dagger}$ and $\hat{\beta}_{\bf k}$, which obey a canonical commutation relation $[\hat{\beta}_{\bf k}, \hat{\beta}_{\bf k'}^{\dagger }]=\delta_{{\bf k},{\bf k'}}$. Their $\theta(N-\hat{n})$-cutoff counterparts are the subspace $\mathcal{H}_N$, which spans only the Fock states with a total occupation $n\in [0,N]$ not larger than the number of particles $N$, and operators $\hat{\beta}_{\bf k}^{'\dagger}=\theta(N-\hat{n})\hat{\beta}_{\bf k}^{\dagger}$, $\hat{\beta}^{'}_{\bf k}=\hat{\beta}_{\bf k}\theta(N-\hat{n})$. They constitute a representation, which is exactly isomorphic to the original one and will be used from now on. In the sense of that isomorphism, one has $\mathcal{H}_N^{(0)}=\mathcal{H}_N$, $\hat{\alpha}_{\bf k}=\hat{\beta}_{\bf k}, \hat{n}_{\bf k}=\hat{\beta}_{\bf k}^{\dagger}\hat{\beta}_{\bf k}$.

    In this representation a standard two-body interaction Hamiltonian \cite{AGD,HohenbergMartin,FetterWalecka,LLV,LL,Anderson1984,Shi,PitString,RevModPhys2004,ProukakisTutorial2008} with a symmetric potential $U({\bf r_1}-{\bf r_2})=U({\bf r_2}-{\bf r_1})$ for the canonical ensemble is equal to
\begin{equation}
H_{N}^{'} =\int \hat{\Psi}^{'\dagger}_N({\bf r_1}) \hat{\Psi}^{'\dagger}_{N-1}({\bf r_2}) U \hat{\Psi}^{'}_{N-1}({\bf r_2}) \hat{\Psi}^{'}_N({\bf r_1}) \frac{d^3 r_1 d^3 r_2}{2},
\label{Hint}
\end{equation}
where $\hat{\Psi}^{'}_N=\hat{\Psi}_N\theta(N-\hat{n})$, $\hat{\Psi}_N({\bf r})=\sqrt{\frac{N-\hat{n}}{V}}+\hat{\beta}_{\bf r}$, $\hat{\beta}_{\bf r}=$ $\frac{1}{\sqrt{V}}\sum_{{\bf k}\neq 0} \hat{\beta}_{\bf k} e^{i{\bf kr}}$, and the zeroth order, ideal-gas Hamiltonian is $H_0 = \sum_{{\bf k}\neq 0} \varepsilon_{\bf k} \hat{n}_{\bf k}$, $\hat{n}_{\bf k}=\hat{\beta}_{\bf k}^{\dagger}\hat{\beta}_{\bf k}$. The theory and calculations are done in terms of the unconstrained excited-states space $\mathcal{H}$ and operators $\hat{\beta}_{\bf k}^{\dagger}$ and $\hat{\beta}_{\bf k}$ by means of the nonpolynomial diagram technique \cite {KochLasPhys2007,KochJMO2007} that allows us to exactly account for all broken symmetry constraints and nonpolynomial functions, like $\theta(N-\hat{n})$ and $\sqrt{N-\hat{n}}$. The latter enter Hamiltonian in Eq. (\ref{Hint}) and are responsible for the so-called constraint nonlinear interaction, that appears in the physical, constrained Hilbert space $\mathcal{H}_N^{(0)}=\mathcal{H}_N$ in addition to two-body interaction in the original, unconstrained, auxiliary Fock space $\mathcal{H}^{(0)}$.

   A primary question to a microscopic theory is a derivation of the equations for the mean value and correlations of the Matsubara annihilation and creation operators \cite{AGD,HohenbergMartin,FetterWalecka,LLV,LL,Anderson1984,Shi}, both the unconstrained ($\tilde{\beta}_{{\bf k}\tau}, \tilde{\bar{\beta}}_{{\bf k}\tau}$) and true ($\tilde{\beta}^{'}_{{\bf k}\tau}=\tilde{\beta}_{{\bf k}\tau}\theta(N-\tilde{n}_{\tau}), \tilde{\bar{\beta}}'_{{\bf k}\tau} =\theta(N-\tilde{n}_{\tau})\tilde{\bar{\beta}}_{{\bf k}\tau}$) ones. Here $\tilde{A}_{\tau}=e^{\tau H}\hat{A}e^{-\tau H}$ for any operator $\hat{A}$, $H=H_0 +H_{N}^{'}$ is a total Hamiltonian, and an imaginary time $\tau \in [0, \frac{1}{T}]$ is related to inverse temperature $\frac{1}{T}$. The definitions for the unconstrained (auxiliary) and true coherent order parameters and Matsubara Green's functions are, respectively,
\begin{equation}
\bar{\beta}_{\bf k}=\langle \hat{\beta}_{\bf k} \rangle , \quad \langle \dots \rangle \equiv  Tr \{\dots e^{-H/T}\}/Tr\{e^{-H/T}\} ,
\label{order}
\end{equation}
\begin{equation}
\bar{\beta}^{'}_{\bf k}= \langle \hat{\beta}^{'}_{\bf k} \rangle_{\mathcal{H}_N}\equiv \frac{\langle \hat{\beta}_{\bf k}\theta(N-\hat{n}) \rangle}{P_N} , \quad P_N = \langle \theta(N-\hat{n}) \rangle ,
\label{order'}
\end{equation}
\begin{equation}
G^{j_2{\bf k_2}\tau_2}_{j_1{\bf k_1}\tau_1}= -\langle T_{\tau} \tilde{\beta}_{j_1{\bf k_1}\tau_1}\tilde{\bar{\beta}}_{j_2{\bf k_2}\tau_2} \rangle ,
\label{Green}
\end{equation}
\begin{equation}
G^{'j_2{\bf k_2}\tau_2}_{j_1{\bf k_1}\tau_1}= -\langle T_{\tau} \tilde{\beta'}_{j_1{\bf k_1}\tau_1}\tilde{\bar{\beta'}}_{j_2{\bf k_2}\tau_2} \rangle /P_N .
\label{Green'}
\end{equation}   
Here $\langle \dots \rangle$ or $\langle \dots \rangle_{\mathcal{H}_N}$ mean the averages over the unconstrained or $\theta(N-\hat{n})$-cutoff excited-states Hilbert spaces $\mathcal{H}$ or ${\mathcal{H}_N}$, respectively. For $\theta(N-\hat{n})$-cutoff operators, like in Eqs. (\ref{order'}) and (\ref{Green'}), they differ only by a factor $P_N$ equal to a cumulative probability of a total occupation of the excited states in the space $\mathcal{H}$ to not exceed the number of particles $N$. $T_{\tau}$ means a standard Matsubara $\tau$-ordering. The indexes $j_{1}=1, 2$ and $j_{2}=1, 2$ enumerate $2\times 2$-matrix of the normal and anomalous Green's functions, since we denote $\tilde{A}_1\equiv \tilde{A}$ and $\tilde{A}_2\equiv \tilde{\bar{A}}$ (a Matsubara-conjugated operator) for any operator $\tilde{A}_{\tau}$.

    Both canonical Green's functions $G$ and $G'$ in Eqs. (\ref{Green})-(\ref{Green'}) are different from the grand-canonical-ensemble ones, employed in a usual Beliaev-Popov theory \cite{AGD,HohenbergMartin,FetterWalecka,Anderson1984,Shi,PitString,RevModPhys2004,ProukakisTutorial2008,LLV,LL} in $\mathcal{H}^{(0)}$. 

$ $

\noindent {\bf 2. Fundamental equations for the unconstrained excitations}

\noindent Exact microscopic equations for the unconstrained coherent order parameter and Green's functions can be derived by means of the nonpolynomial diagram technique \cite{KochLasPhys2007,KochJMO2007} and total irreducible self-energy $\Sigma^{j_2x_2}_{j_1x_1}$, as it is implied for the Dyson-type equations. The result is 
\begin{equation}
\bar{\beta}_{jx}= \check{G}^{(0)}[\check{\Sigma}[\bar{\beta}_{jx}]], \ \check{K}[f_{jx}]\equiv \sum_{j'=1}^2 \int K^{j'x'}_{jx} f_{j'x'} d^4 x' ;
\label{betaEq}
\end{equation}
\begin{equation}
G_{j_1x_1}^{j_2x_2}= -\bar{\beta}_{j_1x_1}\bar{\beta}^{*}_{j_2x_2} + G_{j_1x_1}^{(0)j_2x_2} + \check{G}^{(0)}[\check{\Sigma}[G_{j_1x_1}^{j_2x_2}]], 
\label{GEq}
\end{equation}
where $x=\{{\bf r},\tau\}$, $\int \dots d^4 x \equiv \int_0^{1/T} \int_V \dots d^3 r d\tau$, and an integral operator $\check{K}=\check{\Sigma}$ or $\check{G}^{(0)}$, applied to any function $f_{jx}$, is equal to a convolution of that function $f_{jx}$ over the variables $j, {\bf r}, \tau$ with the self-energy $\Sigma$ or unconstrained zeroth-order Green's function $G^{(0)}$, respectively.

   Using a known for a box, inverse to $\check{G}^{(0)}$ operator \cite{FetterWalecka}, we can rewrite Eqs. (\ref{betaEq})-(\ref{GEq}) in a differential form:
\begin{equation}
\left[ \frac{\hbar \partial}{\partial \tau}+(-1)^j \frac{\hbar^2\nabla_{\bf r}^2}{2M} \right] \bar{\beta}_{jx}= -\sum_{j'=1}^2 \int \Sigma^{j'x'}_{jx} \bar{\beta}_{j'x'} d^4 x' ,
\label{betaEqD}
\end{equation}
$\left[ \frac{\hbar \partial}{\partial \tau_1}+(-1)^{j_1} \frac{\hbar^2\nabla_{\bf r_1}^2}{2M} \right] (G_{j_1x_1}^{j_2x_2} + \bar{\beta}_{j_1x_1}\bar{\beta}^{*}_{j_2x_2})= (-1)^{j_1}\delta_{j_1,j_2}$
\begin{equation}
\times\delta(\tau_1-\tau_2)[\delta({\bf r_1}-{\bf r_2})-\frac{1}{V}] - \sum_{j'=1}^2 \int \Sigma^{j'x'}_{j_1x_1} G_{j'x'}^{j_2x_2} d^4 x' .
\label{GEqD}
\end{equation}

    The interaction effects on the spontaneous symmetry breaking are encoded in the self-energy $\Sigma^{j_1x_1}_{j_2x_2}$. It can be found explicitly from its definition ($\kappa=\{ {\bf k},\tau\}$),
\begin{equation}
-\langle T_{\tau} [\tilde{\beta}_{j_1\kappa_1}, \tilde{H}^{'}_{N\tau_1}]\Delta\tilde{\bar{\beta}}_{j_2\kappa_2} \rangle = \sum_{j=1}^2 \int_0^{1/T} \sum_{{\bf k}\neq 0}\Sigma_{j_1\kappa_1}^{j\kappa}G_{j\kappa}^{j_2\kappa_2} d\tau, 
\label{self-energy}
\end{equation}
with an exact account for ${\bar{\beta}}_{j\kappa}$ if we substitute $\tilde{\beta}_{j\kappa} = \bar{\beta}_{j\kappa}+\Delta\tilde{\beta}_{j\kappa}$ in the l.h.s. $\tilde{H_{N}^{'}}$ and do the first-order, second-order or ladder approximation in interaction, when calculating the l.h.s. average in accord with the main theorem of the diagram technique in the interaction representation over a zeroth-order Hamiltonian $H_0^{(\Delta)} = \sum_{{\bf k}\neq 0} \varepsilon_{\bf k} \Delta\hat{\beta}_{\bf k}^{\dagger}\Delta\hat{\beta}_{\bf k}$. In that way, the nonpolynomial diagram technique \cite{KochLasPhys2007,KochJMO2007} yields the l.h.s. in Eq. (\ref{self-energy}) exactly in a form of its r.h.s.. The calculations are straightforward, but a full analysis is lengthy and will be presented elsewhere. It is based on a distribution of the noncondensate occupation for an ideal gas \cite{PRA2010,PRA2014}, $\rho_n =\int^{\pi}_{-\pi}e^{\phi-iun}\frac{du}{2\pi}, \ \phi= \sum^{\infty}_{m=1}\tilde{\kappa}_m^{(\infty)} \frac{(e^{iu}-1)^m}{m!}$, given for the case $\bar{\beta}_{\bf k}\neq 0$ by the generating cumulants
\begin{equation}
\tilde{\kappa}_m^{(\infty)} =(m-1)! \sum_{{\bf k}\neq 0} [\bar{n}_{\bf k}^{m}+m\bar{n}_{\bf k}^{m-1}|\bar{\beta}_{\bf k}|^2], \bar{n}_{\bf k}= (e^{\frac{\varepsilon_{\bf k}}{T}}-1)^{-1}.
\label{rho}
\end{equation}
The exact Hamiltonian in Eq. (\ref{Hint}) consists of 7 terms:
\begin{equation}
  H_{N}^{'}= \frac{u_0}{2}N(N-1)\theta(N-\hat{n})+V_{1,1}+V_2+V_2^{\dagger}+V_3+V_3^{\dagger}+V_4, 
\label{Hint7}
\end{equation}
\noindent $V_{1,1}= \frac{N-\hat{n}}{V}\theta(N-\hat{n}) \int \int U({\bf r_1}-{\bf r_2})\hat{\beta}_{\bf r_2}^{\dagger}\hat{\beta}_{\bf r_1} d^3 r_1 d^3 r_2,$

\noindent $V_2 = \frac{\hat{Q}}{2V}\theta(N-\hat{n}) \int \int U({\bf r_1}-{\bf r_2})\hat{\beta}_{\bf r_2}\hat{\beta}_{\bf r_1} d^3 r_1 d^3 r_2,$

\noindent $V_3 = \frac{\sqrt{N-\hat{n}}}{\sqrt{V}}\theta(N-\hat{n}) \int \int U({\bf r_1}-{\bf r_2})\hat{\beta}_{\bf r_2}^{\dagger}\hat{\beta}_{\bf r_2}\hat{\beta}_{\bf r_1} d^3 r_1 d^3 r_2,$

\noindent $V_4 = \frac{1}{2}\theta(N-\hat{n}) \int \int (U-u_0)\hat{\beta}_{\bf r_1}^{\dagger}\hat{\beta}_{\bf r_2}^{\dagger}\hat{\beta}_{\bf r_2}\hat{\beta}_{\bf r_1} d^3 r_1 d^3 r_2;$

\noindent $u_{\bf k}=\int_V U({\bf r})e^{-i{\bf kr}}d^3r/V, \ \hat{Q}=\sqrt{(N-\hat{n})(N-1-\hat{n})}$. 

\noindent In the consistent microscopic theory the Heisenberg equation of motion $\partial \tilde{\beta}_x /\partial \tau = [\tilde{\beta}_x ,H]$ contains a commutator
\begin{equation}
[\tilde{\beta}_x , \tilde{H}_{N\tau}^{'}] = (\tilde{H}_{(N-1)\tau}^{'}-\tilde{H}_{N\tau}^{'})\tilde{\beta}_x +\tilde{A} - \int_V \tilde{A}\frac{d^3 r}{V},
\label{commutator}
\end{equation}
$\tilde{A}= \theta(N-1-\tilde{n}_{\tau})\int U({\bf r'}-{\bf r}) \tilde{\bar{\Psi}}_{(N-1)x'} [\tilde{\Psi}_{(N-1)x'}\tilde{\Psi}_{Nx}$ $+ \tilde{\Psi}_{(N-1)x}\tilde{\Psi}_{Nx'}] \frac{d^3 r'}{2},$
with an operator $\tilde{H}_{(N-1)\tau}^{'}-\tilde{H}_{N\tau}^{'}$, which is incorrectly replaced by a c-number chemical potential $\mu$ in the usual approach. The Hamiltonian $H_{N}^{'}$ turns into the usual Beliaev-Popov one if one replaces the operators $\theta(N-\hat{n})$ and $\sqrt{N-\hat{n}}$ by c-numbers. Such replacement is wrong near the critical $\lambda$-point due to very large critical fluctuations, provided by those operators.

$ $

\noindent {\bf 3. Exact equations for the constrained, physical excitations}

\noindent The true order parameter and Green's functions in Eqs. (\ref{order'}) and (\ref{Green'}) can be found from the unconstrained ones:
\begin{equation}
\bar{\beta}^{'}_{j{\bf k}}= \frac{\langle T_{\tau}\hat{S}_{j{\bf k}\tau|_j}[\theta (N-\tilde{n}_\tau)] \rangle}{P_N}, \ \tau|_j \equiv \tau -(-1)^j 0 , \ \bar{j} \equiv 3-j , 
\label{beta'}
\end{equation}
\begin{equation}
G^{'j_2{\bf k_2}\tau_2}_{j_1{\bf k_1}\tau_1}= \frac{\langle T_{\tau}\{ \hat{S}_{\bar{j}_2{\bf k_2}\tau_2|_{\bar{j}_2}}[ \hat{S}_{j_1{\bf k_1}\tau_1|_{j_1}}[\theta]] + \hat{B}^{j_2{\bf k_2}\tau_2|_{\bar{j}_2}}_{j_1{\bf k_1}\tau_1|_{j_1}}[\theta]\} \rangle}{-P_N};
\label{G'}
\end{equation}
$\theta =\theta (N-\tilde{n}_{\tau_1})\theta (N-\tilde{n}_{\tau_2})$. Denoting $J= \{j{\bf k_i}\tau_i \}$, $J|_j= \{j{\bf k_i}\tau_i|_j \}$, we define a partial-contraction superoperator $\hat{S}_J[f]$ or $\hat{B}^{J_2}_{J_1}[f]$, acting on a function $f(\tilde{n}_{\tau'_1},\tilde{n}_{\tau'_2})$, as a linear operator producing from any $\tilde{n}_{\tau'_1}^{m_1}\tilde{n}_{\tau'_2}^{m_2}$ term of the $f$'s Taylor series a sum of all possible operators $R^{m'_1}_{m'_2}\tilde{n}_{\tau'_1}^{m'_1}\tilde{n}_{\tau'_2}^{m'_2}$ with the smaller powers $m'_j \in [0, m_j]$ and coefficients $R^{m'_1}_{m'_2}$, yielding a contribution to $\langle T_{\tau} \Delta\tilde{\beta}_{J}\tilde{n}_{\tau'_1}^{m_1-m'_1}\tilde{n}_{\tau'_2}^{m_2-m'_2}\rangle$ or $\langle T_{\tau} \Delta\tilde{\beta}_{J_1}\Delta\tilde{\bar{\beta}}_{J_2}\tilde{n}_{\tau'_1}^{m_1-m'_1}\tilde{n}_{\tau'_2}^{m_2-m'_2}\rangle$, respectively, from all possible connected 1D diagrams, which start from an external operator $\Delta\tilde{\beta}_{J}$ or $\Delta\tilde{\beta}_{J_1}$ and via elementary contractions \cite{FetterWalecka} $\langle T_{\tau} \Delta\tilde{\beta}_{J_1}\Delta\tilde{\bar{\beta}}_{J_2}\rangle \equiv -g^{J_2}_{J_1}=-G^{J_2}_{J_1}-\bar{\beta}_{J_1}\bar{\beta}^*_{J_2}$, connecting vertices $\tilde{n}_{{\bf k_i}\tau'_i}=(\Delta\tilde{\bar{\beta}}_{{\bf k_i}\tau'_i}+\bar{\beta}^*_{{\bf k_i}\tau'_i})(\Delta\tilde{\beta}_{{\bf k_i}\tau'_i}+\bar{\beta}_{{\bf k_i}\tau'_i})$, go to the end internal operator $\Delta\tilde{\bar{\beta}}_{{\bf k_i}\tau'_i}$ or $\Delta\tilde{\beta}_{{\bf k_i}\tau'_i}$ (for $\hat{S}$) or end external operator $\Delta\tilde{\bar{\beta}}_{J_2}$ (for $\hat{B}$) in accord with a Wick's theorem. The $\hat{S}_{J}[f]$ includes, in addition, $\bar{\beta}_{j{\bf k_i}}f$.

    The point is that there are no closed equations for the true, constrained order parameter and Green's functions, but we find the exact closed difference (recurrence) equations for the basis one- and two-contraction superoperators $\hat{s}_{J}(p,q)= T_{\tau} \hat{S}_{J|_j}[f(p,q)]$ and $\hat{b}^{J_2}_{J_1}(p,q)= T_{\tau} \hat{B}^{J_2|_{\bar{j}_2}}_{J_1|_{j_1}}[f(p,q)]$ for any $f(p,q)=f(\tilde{n}_{\tau_1}+p,\tilde{n}_{\tau_2}+q)$:
\begin{equation}
\hat{s}_{J}(p,q)= \bar{\beta}_{j{\bf k_i}}f(p,q)- g_{J|_{j}}^{J'} \Delta_{p}^{\delta_1^{J'}} \Delta_{q}^{\delta_2^{J'}} \hat{s}_{J'}(p,q),
\label{1-contraction}
\end{equation}
\begin{equation}
\hat{b}^{J_2}_{J_1}(p,q)= -g^{J_2|_{\bar{j}_2}}_{J_1|_{j_1}}f(p,q) + g_{J_1|_{j_1}}^{J'} g_{J'}^{J_2|_{\bar{j}_2}} \Delta_{p}^{\delta_1^{J'}} \Delta_{q}^{\delta_2^{J'}}f(p,q) 
\label{2-contraction}
\end{equation}

$+g_{J_1|_{j_1}}^{J'_1} g_{J'_2}^{J_2|_{\bar{j}_2}} \Delta_{p}^{\delta_1^{J'_1}+\delta_1^{J'_2}} \Delta_{q}^{\delta_2^{J'_1}+\delta_2^{J'_2}} \hat{b}^{J'_2}_{J'_1}(p,q).$

   Here $j=1,2$ and $i=1,2$, $\delta^{J'}_{i}\equiv \delta_{i',i}$, $\Delta_p$ and $\Delta_q$ are the partial difference operators \cite{DE-Elaydi} ($\Delta_p f(p,q)=f(p,q)-f(p-1,q)$ and $\Delta_q f(p,q)=f(p,q)-f(p,q-1)$), $\Delta_p^0 \equiv \Delta_q^0 \equiv 1$, and we assume Einstein's summation over the repeated indexes $J', J'_1, J'_2$. The sums run over $j'=1,2$, $i'=1,2$, ${\bf k'_{i'}}\neq 0$ for $J'$ and similarly for $J'_1, J'_2$. Eqs. (\ref{1-contraction}) and (\ref{2-contraction}) are the linear systems of the integral equations over the wave vector variables and discrete (recurrence) equations over variables $p$ and $q$ with the well-known methods of solution \cite{PDE-Cheng,DE-Agarwal,DE-Elaydi}, e.g., via Z-transform (discrete analog of Laplace transform) or a characteristic function (solution for $f=\exp(iu_1\tilde{n}_{\tau_1}+iu_2\tilde{n}_{\tau_2})$ with a subsequent Fourier transform). The superoperators in Eqs. (\ref{beta'})-(\ref{G'}) are given by those solutions at $p=q=0$. Finally, their average in Eqs. (\ref{beta'})-(\ref{G'}) amounts to the averages like $\langle f(\tilde{n}_{\tau_1},\tilde{n}_{\tau_2}) \rangle$, that is reduced to calculation of a joint distribution $\rho_{n_1,n_2}$ of the noncommuting operators $\tilde{n}_{\tau_1}$ and $\tilde{n}_{\tau_2}$. The latter can be done similar to \cite{PRA2010,PRA2014} via its characteristic function $\langle T_{\tau} \exp(iu_1\tilde{n}_{\tau_1}+iu_2\tilde{n}_{\tau_2}) \rangle$, which is equal to
\begin{equation}
\Theta= \prod_{{\bf k}\neq 0} \frac{e^{\frac{|\bar{\beta}_{\bf k}|^2 [(z_{\bf k}-1)(z_1z_2-1)-(z_{\bf k}-e^{\tau \varepsilon_{\bf k}})(1-e^{-\tau \varepsilon_{\bf k}})(z_1-1)(z_2-1)]}{z_{\bf k}-z_1z_2}}}{[(z_{\bf k}-z_1z_2)/(z_{\bf k}-1)]}
\label{2Dchar-function}
\end{equation}
for any $\bar{\beta}_{\bf k}$ in the zeroth order in interaction; $\tau=\tau_1-\tau_2, z_{\bf k}=e^{\frac{\varepsilon_{\bf k}}{T}}, z_j =e^{iu_j}, j=1,2$. The result in Eq. (\ref{2Dchar-function}), together with its particular case for $u_2=0$ in Eq. (\ref{rho}), provides a basis for a perturbation analysis. The macroscopic wave function (the coherent order parameter) of the excitations' condensate $\bar{\beta}_{\bf r}\neq 0$, entering Eqs. (\ref{GEq}), (\ref{GEqD}), and (\ref{beta'})-(\ref{2-contraction}), makes the quasiparticles stable and can be found from Eqs. (\ref{betaEq}) or (\ref{betaEqD}). 

     Note that within the particular case of the homogeneous BEC in a box with the periodic boundary conditions the usually used Beliaev-Popov equations describe BEC as a smooth change of stable quasiparticles, dressed by a Bogoliubov coupling via a condensate occupation and possessing the inter-mode correlations (the anomalous Green's functions) without any own coherence. In that picture, a coherent superfluid flow is generated by some external sources or boundary motion, but an appearance of the coherent macroscopic order parameter $\bar{\beta}^{'}_{\bf r}$ is excluded by the Hugenholtz-Pines ($T=0$) or Hohenberg-Martin ($T\neq 0$) condition \cite{HohenbergMartin} $(\Sigma^{1{\bf k}}_{1{\bf k}} -\Sigma^{1{\bf k}}_{2{\bf -k}})|_{{\bf k}=0} =\mu$, that ensures a stable gapless spectrum of quasiparticles. That usual mechanism of BEC is similar to the one in an ideal gas, being just its dressed-by-interaction version. It differs from the outlined general mechanism, allowing an instability of restructuring of the condensate by $\bar{\beta}^{'}_{\bf r}\neq 0$, that is a symmetry breaking (cf. a density wave in a superfluid \cite{Anderson1984}). A detailed analysis of Eqs. (\ref{betaEq})-(\ref{GEqD}), (\ref{beta'})-(\ref{2-contraction}) will be given elsewhere.

$ $

\noindent {\bf 4. Discussion} 

\noindent The derived fundamental equations (\ref{betaEq})-(\ref{GEqD}), (\ref{beta'})-(\ref{2-contraction}) consistently account for the critical fluctuations as well as for the effects of the inter-particle interaction on the condensate and quasiparticles. These equations have perfectly canonical and proper matrix structure. The unconstrained order parameter $\bar{\beta}_J$ and Green's function $G^{J_2}_{J_1}$ form the wavevector-dependent 4-component vector and $4\times 4$-matrix, respectively, since the combined index $J= \{j{\bf k_i}\tau_i \}$ includes the indexes $j=1,2$ and $i=1,2$, each with two values. The $\bar{\beta}_J$ and $G^{J_2}_{J_1}$ are the solutions of the canonical Dyson-type equations (\ref{betaEq})-(\ref{GEqD}), which are the integral equations over the wave vector variable, and are determined by the $4\times 4$-matrices of the free propagator $G^{(0)J_2}_{J_1}$ and self-energy $\Sigma^{J_2}_{J_1}$, defined by the two-body interaction in Eq. (\ref{self-energy}). In their turn, these solutions $\bar{\beta}_J$ and $G^{J_2}_{J_1}$ enter the inhomogeneous linear integral matrix equations (\ref{1-contraction})-(\ref{2-contraction}) for the basis one- and two-contraction superoperators $\hat{s}_{J}(p,q)$ (4-vector) and $\hat{b}^{J_2}_{J_1}(p,q)$ ($4\times 4$-matrix) via the source terms and coefficients. The coefficients are given by the Green's function $4\times 4$-matrix $g^{J_2}_{J_1}=G^{J_2}_{J_1}+\bar{\beta}_{J_1}\bar{\beta}^*_{J_2}$ (cf. a usual Nambu's $2\times 2$-matrix of the normal and anomalous Green's functions). In addition to being matrix and integral, the latter equations (\ref{1-contraction})-(\ref{2-contraction}) are the partial difference equations (a discrete analog of the partial differential equations) over the variables $p$ and $q$. The difference equations are required by a discrete, quantum nature of the excitations.

   The revealed nontrivial universal structure of these fundamental equations is uniquely prescribed by the microscopic physics of the critical phenomena because these equations are not the model, approximate, or phenomenological ones, but the exact equations. It is immediate to generalize them to any trapping potentials and boundary conditions. They open a way to solve the long-standing problem of the BEC and other phase transitions \cite{AGD,HohenbergMartin,FetterWalecka,Anderson1984,Shi,PitString,RevModPhys2004,ProukakisTutorial2008,LLV,LL,PatPokr,Fisher1986}, including a restricted canonical ensemble problem \cite{HohenbergMartin}, and describe numerous modern laboratory and numerical experiments on the critical phenomena in BEC of the mesoscopic systems \cite{BrownTwissOnBECThresholdNaturePhys2012,Hadzibabic2013NaturePhys,Hadzibabic2013,EsslingerCriticalBECScience2007,TwinAtomBeamNaturePhys2011,Dalibard2012NaturePhys,Dalibard2010,RaizenTrapControl,Blume,BECinterferometerOnChip2013,EntanglementOnChipNature2010,chipBECNature2001,HeDroplets,Reppy}. 

   A principle difference of these microscopic equations from the usual Gross-Pitaevskii and Beliaev-Popov equations comes from a presence of the operator functions $\theta(N-\hat{n})$ and $\sqrt{N-\hat{n}}$ in the true Hamiltonian (\ref{Hint}) and the symmetry-constrained nature of the actual excitation operators $\hat{\beta}^{'}_{\bf k}=\hat{\beta}_{\bf k}\theta(N-\hat{n})$, which determine the true order parameter and Green's functions in Eqs. (\ref{order'}) and (\ref{Green'}). Far outside the critical region, at low enough temperatures, when the ordered phase is fully formed and the order parameter is much larger than its fluctuations, a mean-field approximation, replacing those operator functions by the c-numbers $\theta(N-\hat{n})\approx 1, \sqrt{N-\hat{n}}\approx \bar{n}_0$, becomes valid and the exact microscopic equations asymptotically turn into the usual Gross-Pitaevskii and Beliaev-Popov ones.

   The main point is that the microscopic theory is valid both inside and outside the critical region. In particular, it is capable of a microscopic calculation of the critical exponents and other parameters of the BEC phase transition near the $\lambda$-point for an actual, canonical-ensemble mesoscopic system, as opposed to a grand-canonical-ensemble bulk-limit model, usually studied in the phenomenological renormalization-group theory \cite{RevModPhys2004,LLV,LL,PatPokr,Fisher1986,Goldenfeld,Vicari2002}.

    There is another known limit of this microscopic theory. Namely, for a vanishing interaction, when $H_{N}^{'} \equiv 0$ and $\bar{\beta}_{\bf k} \equiv 0$ in Eqs. (\ref{rho}) and (\ref{2Dchar-function}), it coincides with a solution for the critical fluctuations in the BEC of a mesoscopic ideal gas \cite{PRA2010,PRA2014}, which is its zeroth-order approximation.
    
     In short, all these advances come from the correct account, first, of the Noether's symmetry constraints in the many-body Hilbert space and, second, of the related properties of the true excitations in a mesoscopic system.

     Finally, it is straightforward to extend the general method and structure of this microscopic theory of the critical phenomena to other phase transitions in various fields of physics, including the physics of condensed matter and quantum fields. This Letter aims to bring attention to a remarkable opportunity to solve that open problem, which is common to so many physical systems.

   In summary, we find a microscopic theory of a phase transition in a critical region, a critical-region extension of the Gross-Pitaevskii and Beliaev-Popov equations, an exact Hamiltonian for the Bose-Einstein condensation in a mesoscopic system, and the exact recurrence equations for the basis contraction superoperators.

\end{document}